\documentclass[onecolumn,final,aps,prb,showpacs,amsmath,amssymb,amsfonts,floatfix,superscriptaddress,nobibnotes]{revtex4-1}

\usepackage{amsmath,amssymb,amsfonts}
\usepackage{color}
\usepackage[colorlinks,breaklinks,bookmarks=true,citecolor=blue,linkcolor=red,urlcolor=blue]{hyperref}
\definecolor{darkred}{rgb}{0.7,0.0,0}

\usepackage{graphicx}
\usepackage{multirow}
\usepackage{epsfig}
\usepackage{braket}
\usepackage{lineno}
\def\FGT{$\mathrm{Fe_3GeTe_2}$ }

\begin{document}
\title{$\mathrm{Fe_3GeTe_2}$: A site-differentiated Hund metal}

	\author{Taek Jung Kim}
	\email{tj.kim0530@gmail.com}	
	\affiliation{Department of Physics, KAIST, Daejeon 34141, Republic of Korea}

	\author{Siheon Ryee}
	\email{sryee@physnet.uni-hamburg.de}
	\affiliation{Department of Physics, KAIST, Daejeon 34141, Republic of Korea}
	\affiliation{I. Institute of Theoretical Physics, University of Hamburg, Notkestrasse 9-11, Hamburg 22607, Germany}
	
	\author{Myung Joon Han}
	\email{mj.han@kaist.ac.kr (corresponding author)}
	\affiliation{Department of Physics, KAIST, Daejeon 34141, Republic of Korea}

	\date{\today}

	\newenvironment{Acknowledgements}{\section*{Acknowledgements}\appendix}{}
	
	\maketitle

%\linenumbers %%%%%%%%%%%%%%%%%%%%%%%%%%%  MJHMJH

    {\bf
    Magnetism in two-dimensional (2D) van der Waals (vdW) materials has lately attracted considerable attention from the point of view of both fundamental science and device applications. Obviously, establishing the detailed and solid understanding of their magnetism is the key first step toward various applications. Although $\mathrm{Fe_3GeTe_2}$ is a representative ferromagnetic (FM) metal in this family, many aspects of its magnetic and electronic behaviors still remain elusive. Here, we report our new finding that $\mathrm{Fe_3GeTe_2}$ is a special type of correlated metal known as `Hund metal'. Furthermore, we demonstrate that Hund metallicity in this material is quite unique by exhibiting remarkable site-dependence of Hund correlation strength, hereby dubbed `site-differentiated Hund metal'. Within this new picture, many of previous experiments can be clearly understood including the ones that were seemingly contradictory to one another.
    } 

\section*{Introduction}
$\mathrm{Fe_3GeTe_2}$ has been a focus of recent surge of 2D magnetic material research \cite{burch_magnetism_2018,gong_two-dimensional_2019,mak_probing_2019,gibertini_magnetic_2019,deiseroth_fe3gete2_2006,chen_magnetic_2013,fei_2018,jang_origin_2020,zhang_emergence_2018,deng_gate-tunable_2018,wang_above_2020,kim_large_2018}. On one hand, it displays many fascinating properties such as high  critical temperature ($T_\mathrm{c} \simeq 220$~K) \cite{chen_magnetic_2013,jang_origin_2020}, heavy fermion behavior \cite{zhang_emergence_2018}, and the relative ease of exfoliation \cite{deiseroth_fe3gete2_2006,fei_2018}. Its great potential has also been highlighted by recent experiments showing that $T_\mathrm{c}$ can be increased even up to room temperature \cite{deng_gate-tunable_2018,wang_above_2020} and that the large anomalous Hall current is originated from the unique band topology \cite{kim_large_2018}.
On the other hand,  its electronic and magnetic properties are far from being clearly understood. The very basic picture for the magnetic moment is still under debate \cite{zhu_electronic_2016,zhang_emergence_2018,xu_signature_2020}. 
The measured quasiparticle masses are not consistent with each other and exhibit an order of magnitude difference depending on experimental probes  \cite{zhu_electronic_2016,kim_large_2018,zhang_emergence_2018,xu_signature_2020}. These puzzles are posing a challenge to the current theory of this material.

In this work, we suggest a new physical picture for $\mathrm{Fe_3GeTe_2}$. We first provide convincing evidences that \FGT is a `Hund metal'.
The term Hund metal was coined to refer to an intriguing type of correlated (not necessarily magnetic) metals\cite{haule_coherenceincoherence_2009,yin_kinetic_2011,de_medici_hunds_2011,de_medici_janus-faced_2011,georges_strong_2013} in which Hund coupling $J_\mathrm{H}$, rather than Hubbard $U$, plays the main role in determining the electronic properties\cite{georges_strong_2013}. Hund metals can host various interesting phenomena such as spin freezing \cite{werner_spin_2008,hoshino_superconductivity_2015}, spin-orbital separation\cite{okada_singlet_1973,yin_fractional_2012,stadler_dynamical_2015,horvat_low-energy_2016,deng_signatures_2019}, orbital differentiation  \cite{de_medici_selective_2014,kostin_imaging_2018}, and unconventional superconductivity \cite{hoshino_superconductivity_2015,lee_pairing_2018}. This concept has been providing a compelling view for many of multiorbital systems\cite{werner_spin_2008,haule_coherenceincoherence_2009,yin_magnetism_2011,mravlje_coherence-incoherence_2011,bascones_orbital_2012,georges_strong_2013,lanata_orbital_2013,hoshino_superconductivity_2015,khajetoorians_2015,fanfarillo_1,mravlje_thermopower_2016,de_medici_hunds_2017,lee_pairing_2018,isidori,ryee_nonlocal_2020,kugler_strongly_2020, chen_unconventional_2020,ryee_2021,fanfarillo_2,karp_2020}, most prominently for iron-based superconductors\cite{yin_magnetism_2011,haule_coherenceincoherence_2009,georges_strong_2013, bascones_orbital_2012,lanata_orbital_2013,de_medici_hunds_2017,lee_pairing_2018,watzenbock_2020}, ruthenates\cite{werner_spin_2008,hoshino_superconductivity_2015,mravlje_coherence-incoherence_2011,mravlje_thermopower_2016,kugler_strongly_2020}, and presumably also for recently discovered nickelate superconductors \cite{wang_hunds_2020,kang_infinite-layer_2021}. Our analysis demonstrates that $\mathrm{Fe_3GeTe_2}$ is a new member of this family.

\begin{figure}[t]
	\includegraphics[width=8.4cm]{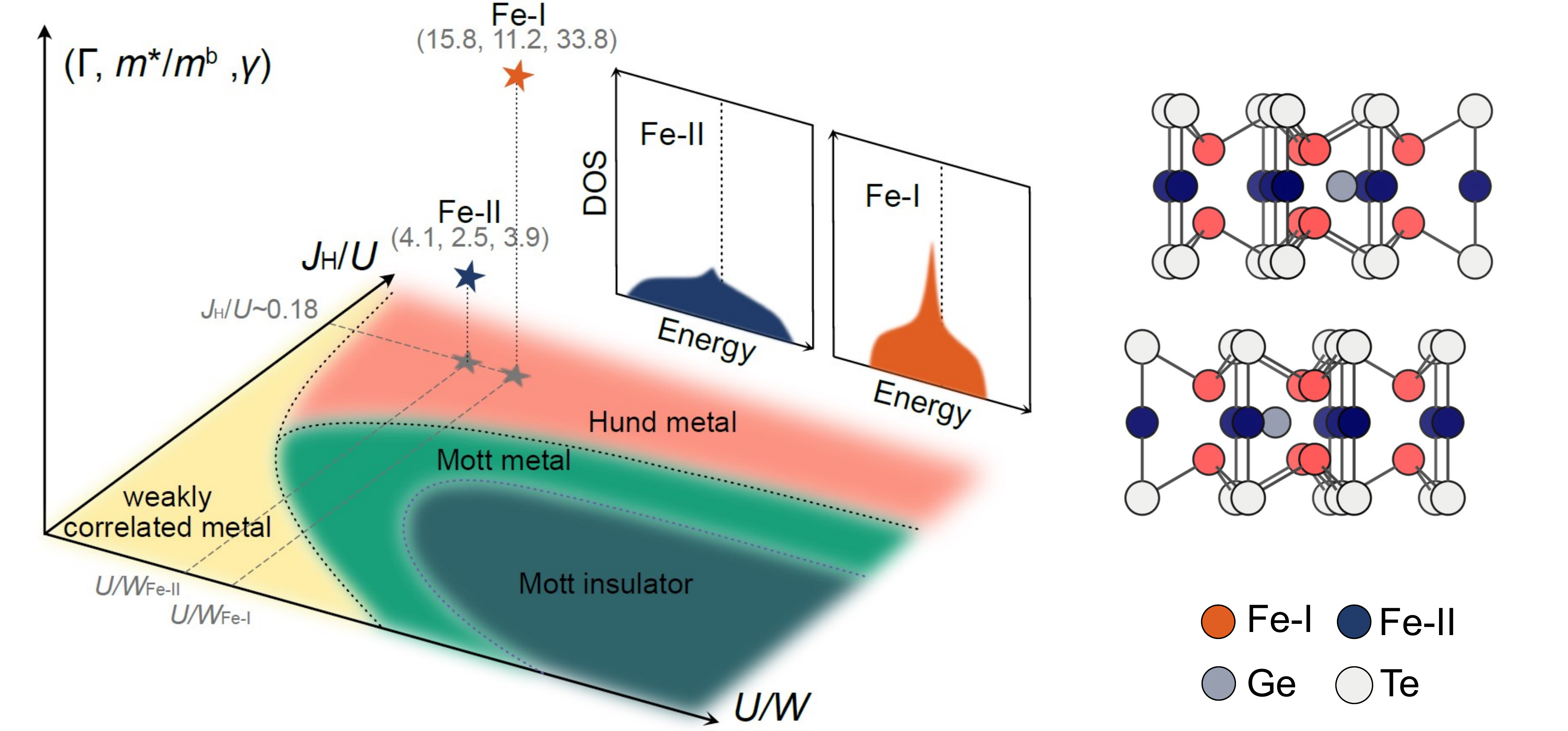}
	\caption{ {\bf Overview figure.} {Left panel: A schematic illustration of `site-differentiated Hund metal' physics in $\mathrm{Fe_3GeTe_2}$. Colored areas highlight different regimes or phases; Mott insulator (dark green; i.e., whose charge gap is opened due to Hubbard $U$), Mott metal (light green; i.e., whose correlated metallic behavior is governed by $U$), weakly correlated metal (yellow; electronic correlation is weak enough), and Hund metal (red; i.e., whose correlated metallic behavior is governed by Hund $J_\mathrm{H}$). For more details to define each region of phase diagram, we refer to Refs.~\onlinecite{ryee_2021,strand_valence-skipping_2014,fanfarillo_1}.
			While both Fe sites fall into the Hund metal regime, they are clearly differentiated in terms of electron scattering rate $\Gamma$, quasiparticle effective mass $m^*/m^{\rm b}$, and Sommerfeld coefficient $\gamma$ (as depicted along $z$-axis) which are attributed to the electronic structure details (see the schematic DOS). In parentheses shown are the calculated spin-orbital-averaged values of ($\Gamma$, $m^*/m^{\rm b}$, $\gamma$)  taken at T=100~K. The unit of $\Gamma$ is meV. The $x$- and $y$-axis corresponds to $J_\mathrm{H}/U$ and $U/W$, respectively. $U=$5.0 and $J=$0.9~eV ($J_\mathrm{H}/U=$0.18) were used for both Fe-I and Fe-II; see the method section for more detailed discussion. The effective bandwidth $W_\mathrm{Fe-I}$ for Fe-I is found to be smaller than that of Fe-II through the analysis of hybridization function (see main text). A rough estimation of their difference  based on the integrated DOS yields $\sim$1.6~eV  (see main text). Right panel: Crystal structure of $\mathrm{Fe_3GeTe_2}$.}
		\label{fig1}}
\end{figure}

Further, we show that Hund metallicity of this material is distinctive from other known Hund metals by exhibiting the remarkable site dependence of correlation strengths; dubbed `site-differentiated' Hund metal. (see Fig.~\ref{fig1}). This intriguing site dependence is originated from the microscopic details of underlying band structure. The suggested new picture is an important key to understand experiments including the ones that are seemingly contradictory to each other.

\section*{Results and Discussion}

\subsection*{$\mathrm{Fe_3GeTe_2}$ is a Hund metal}
\FGT is a vdW material with a layered $\mathrm{Fe_{3}Ge}$ substructure sandwiched by two Te layers. Five Fe-$3d$ levels of both Fe-I and Fe-II are split into three groups, namely $d_{\rm z^2}, d_{\rm x^2-y^2/xy}$ and $d_{\rm xz/yz}$. More details can be found in Supplementary Note~1 of Supplementary Information (SI).

%\vspace{3mm}\textbf{\\ \FGT is a Hund metal } 

We firstly demonstrate that all the key features defining Hund metal are well identified in \FGT via density functional theory plus dynamical mean-field theory (DFT+DMFT) calculations (see Methods). In order to concentrate on the intrinsic nature of correlation (separately from magnetism), we suppress magnetic ordering unless otherwise specified. The result of $U=5$ and $J_\mathrm{H}=0.9$~eV will be presented as our main parameter choice. Computation methods including the parameter dependence and other details are given in Methods.

First, we examine so-called two-faced or `Janus' effect which represents the physics governed by $J_\mathrm{H}$ distinctively from that of Mott correlation $U$ \cite{de_medici_janus-faced_2011,georges_strong_2013}. The straightforward way is to see the simultaneous increase of both $U_\mathrm{c}$ (the critical value of $U$ at which the metal-to-Mott-insulator transition occurs) and `correlation strength' as measured typically by the inverse of quasiparticle weight $Z$  (or equivalently, the quasiparticle scattering rate $\Gamma$) as $J_\mathrm{H}$ increases \cite{de_medici_janus-faced_2011,georges_strong_2013}.

Let us uncover the `first face' of this effect, namely, the $U_\mathrm{c}$ increase by $J_\mathrm{H}$. To avoid the demanding computations scanning the large parameter space, here we simply examine the atomic gap $\Delta_{\rm at}\equiv E_{N+1}+E_{N-1}-2E_{N}$ ($E_{N}$: the atomic ground state energy of $N$-electron subspace). According to the original Mott-Hubbard picture, the gap opening is associated with $\Delta_{\rm at} > W$ ($W$: bandwidth), and the system gets closer to Mott transition as $\Delta_{\rm at}$ increases. This analysis, albeit simple, is well justified by the observations that the actual DMFT phase diagrams are consistent with the behavior of $\Delta_{\rm at}$ in the large $J_\mathrm{H}$ regime \cite{georges_strong_2013,ryee_2021}.
We found that $\Delta_{\rm at}\approx U-1.52J_\mathrm{H}$ from a five-degenerate-orbital model with 6 electrons, which guarantees that $\partial \Delta_{\rm at} / \partial J_\mathrm{H} <0$ (see the inset of Fig.~\ref{fig2}a). Note that $J_\mathrm{H}$ drives the system away from being  a Mott insulator; hereby the Janus' first face is unveiled.

To see its `second face', the calculated $\Gamma$ is presented in Fig.~\ref{fig2}a. Here, $\Gamma_l=Z_l\mathrm{Im}\Sigma_l(\mathrm{i}\omega_n)|_{\omega_n \rightarrow 0^+}$ where $Z_l$ is the quasiparticle weight and $\Sigma_l(\mathrm{i}\omega_n)$ the local self-energy on the imaginary frequency ($\omega_n$) axis for a given orbital $l$ (see Supplementary Note~2 of SI for additional discussion). It is clear that the $\Gamma$ is enhanced as $J_\mathrm{H}$ increases. Also, the scattering rate is strongly orbital-dependent as depicted by different symbols, and the differences become more pronounced as $J_\mathrm{H}$ increases. This feature, often called orbital differentiation, has been regarded as one of the key pieces for identifying iron-based superconductors as a Hund metal \cite{de_medici_selective_2014,kostin_imaging_2018}.

Having evidenced Janus behavior, we now turn to another defining feature of Hund metallicity, namely, spin-orbital separation \cite{okada_singlet_1973,yin_fractional_2012,stadler_dynamical_2015,horvat_low-energy_2016,deng_signatures_2019}. 
	As discussed in previous works, the electronic behavior of Hund metals follows local-atomic and Landau-quasiparticle picture in the high- and low-enough energy region, respectively, due to the gradual development of screenings. 
	Deng et al., suggested that the much higher onset screening temperature for orbital degree of freedom ($T_{\rm onset}^{\rm orb}$) than that of spin ($T_{\rm onset}^{\rm spin}$) is a key signature of Hund metals; {\it i.e,} $T_{\rm onset}^{\rm orb} \gg T_{\rm onset}^{\rm spin}$
	 \cite{deng_signatures_2019}.
Following the literature, we tracked local susceptibilities up to high enough temperature.
Fig.~\ref{fig2}b shows the calculated local spin and orbital susceptibilities multiplied by $k_\mathrm{B}T$ ($k_\mathrm{B}$: Boltzmann constant). The spin susceptibility  with no prefactor $k_\mathrm{B}T$ is presented in Fig.~\ref{fig2}c. The onset screening temperatures for spin (below which $k_\mathrm{B}T\chi^{\rm spin}$ starts to deviate from the Curie law) are indicated by the vertical orange arrows; $T^{\rm spin}_{\rm onset} \approx 3000$~K and 9000~K for Fe-I and Fe-II, respectively. Importantly, the onset temperatures for orbital screenings are much higher; $T^{\rm orb}_{\rm onset} > 20000$~K for both Fe-I and Fe-II; see the empty symbols. The well separated $T_{\rm onset}^{\rm spin}$ and $T_{\rm onset}^{\rm orb}$ provide further evidence that \FGT is a Hund metal.

We also investigated the long-time correlators,
$C^\mathrm{spin/orb}(\tau=1/2k_{\rm B}T)$ ($\tau$: imaginary time); see Supplementary Note~3 of SI for its definition and the related discussion.
These quantities were recently used as another useful proxy to measure the degree of spin-orbital separation \cite{ryee_2021}. The results also clearly show that \FGT is well classified as a Hund metal.

{We finally note that the ferromagnetic transition temperature ($T_\mathrm{c} \simeq220$~K) of $\mathrm{Fe_3GeTe_2}$ is much lower than the characteristic temperature scales of $T_{\rm onset}$ for spin and orbital. It indicates that the Hund metal character is developed well above the ferromagnetic transition temperature and the magnetic order itself is not immediately relevant to Hund metallicity.}

\subsection*{Site-differentiated Hund metallicity}

\begin{figure*}[t]
	\includegraphics[width=17.5cm]{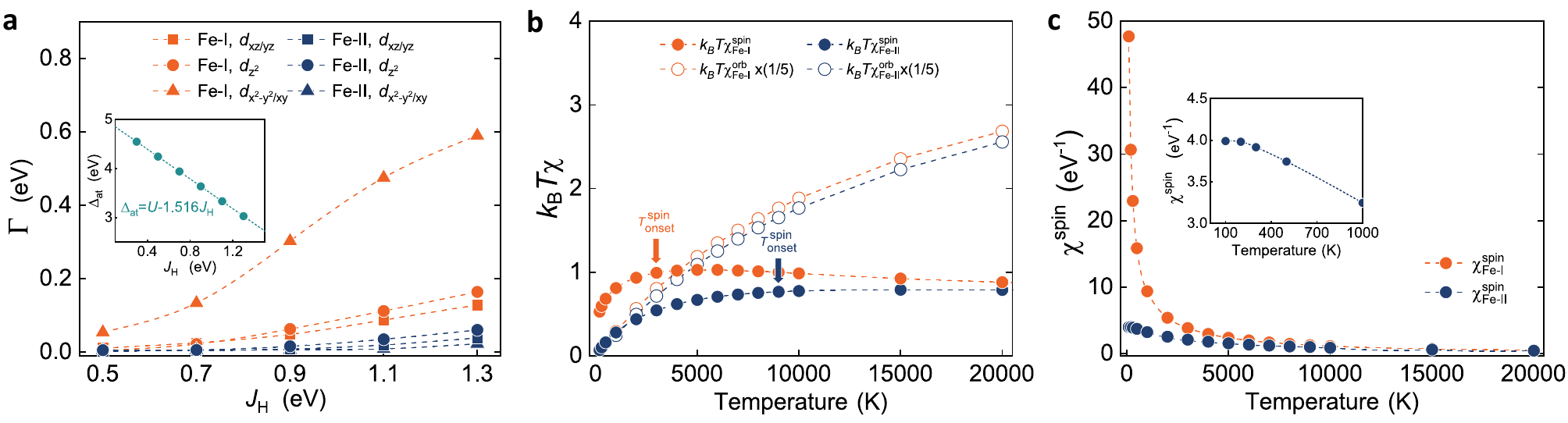}
	\caption{ {\bf Quasiparticle scattering rate $\Gamma$, local susceptibilities multiplied by temperature $k_{\rm B}T\chi$ and the spin susceptibilities with no factor $k_{\rm B}T$. } $\bf{a}$, The $J_{\rm H}$ dependence of quasiparticle scattering rate $\Gamma$ for each orbital in Fe-I (orange) and Fe-II (blue) at $T=300$~K. Inset in ($\bf{a}$) shows the $J_{\rm H}$ dependence of $\Delta_{\rm at}$ for a degenerate five-orbital model with six electrons. $\bf{b,c}$, Temperature dependence of the susceptibilities ($\bf{b}$) $k_{\rm B}T\chi^\mathrm{spin/orb}$ and ($\bf{c}$) $\chi^\mathrm{spin}$ of Fe-I (orange) and Fe-II (blue). The filled and open symbols refer to the spin and orbital susceptibility, respectively. Two arrows in ($\bf{b}$) indicate the onset temperature of spin screening below which $\chi^\mathrm{spin}$ deviates from the Curie law of free local moment. The inset in ($\bf{c}$) is an enlarged view of $\chi^{\rm spin}_{\rm Fe-II}$ in the low temperature region. \label{fig2}}
\end{figure*}

Hund metallicity in \FGT is unique in that its Hund physics is clearly differentiated at two sites. Note that the scattering rate $\Gamma$ is an order-of-magnitude larger in Fe-I. $\Gamma$ of Fe-I $d_{\rm x^2-y^2/xy}$ (orange triangles in Fig.~\ref{fig2}a) reaches 0.6 at $J_\mathrm{H}=1.3$~eV whereas that of Fe-II $d_{\rm z^2}$ only 0.05 (blue circles) for example. The orbital-dependent correlation is also more pronounced in Fe-I (Fig.~\ref{fig2}a).

$\chi^{\rm spin}$ also shows remarkable site-dependence as presented in Fig.~\ref{fig2}c: As temperature is lowered, $\chi^{\rm spin}$ of Fe-I (orange line) well follows $\sim 1/T$ behavior in sharp contrast to Fe-II (blue line) which almost saturates at $\sim$100~K (see inset of Fig.~\ref{fig2}c). It implies that, below this temperature, the spin moments of Fe-II tend to be itinerant. For Fe-I, on the other hand, we could not find any indication of susceptibility saturation down to the lowest temperature we reached ($\sim 100$~K), indicative of the persistent local moment character. See also Supplementary Note~3 of SI for further analysis based on long-time correlators.

{What is the origin of this site-dependent Hund physics? First of all, we note that the $d$-orbital occupations are almost the same in both sites; the calculated electron occupancy is $6.431 \pm 0.001$ at $U=0$ and $6.18 \pm 0.05$ at $U=5$ eV ($-/+$ corresponding to Fe-I/Fe-II). {It is  largely attributed to the strong covalency in this chalcogenide material.} Although the electron number depends on the charge-counting method, this level of coincidence can effectively exclude the possibility of the site-differentiated Hund physics being originated simply from the different charge status or valence. Rather, it indicates the more subtle electronic origins.  Second, the calculated density of states (DOS) does not show significant difference between Fe-I and Fe-II other than the more pronounced van Hove singularity (vHS) for Fe-I $d_{\rm x2-y2/xy}$ ({see Supplementary Note~4 of SI}).}

Highly useful insights are obtained from the calculated `one-shot' hybridization function $\Delta(\mathrm{i}\omega_n)$ for each orbital. For its definition and the related figures, refer to Sec.~S4 of SI. $-{\rm Im}\Delta(\mathrm{i}\omega_n)$ at high energy (which approximately scales as the square of the bandwidth) is smaller for Fe-I than Fe-II (see Supplementary Fig.~4a in SI). It shows that the effective bandwidth for Fe-I is smaller than Fe-II, indicative of the stronger correlation at Fe-I. This feature of smaller bandwidth for Fe-I is also seen in the integrated DOS analysis (see Supplementary Fig.~4b 
	%from which one may also want to estimate the bandwidth difference
	). It is therefore consistent with the larger $\Gamma$ (discussed above; see Fig.~\ref{fig2}a) and the greater mass of Fe-I (to be discussed below; see Fig.~\ref{fig3}d). Also, in the low-frequency regime, $-{\rm Im}\Delta(\mathrm{i}\omega_n)$ of Fe-I $d_{\rm x2-y2/xy}$ is smallest and exhibits a downturn as $\omega_n \rightarrow 0$. Note that $-{\rm Im}\Delta(\mathrm{i}\omega_n)|_{\omega_n \rightarrow 0} \sim 1/\{ (\pi D(E_\mathrm{f}))\}$. Here $ D(E_\mathrm{f})$ is the DOS at the Fermi level $E_\mathrm{f}$. Thus, this downturn of $-{\rm Im}\Delta(\mathrm{i}\omega_n)$ is most likely due to the well-developed vHS (i.e., the divergence of DOS), which in turn suppresses the low-energy effective hopping processes \cite{mravlje_coherence-incoherence_2011}. That is, electron scattering and mass enhancement of Fe-I $d_{\rm x2-y2/xy}$ are further boosted by vHS.

\subsection*{Understanding the experiments}

\begin{figure*}[th]
	\includegraphics[width=15cm]{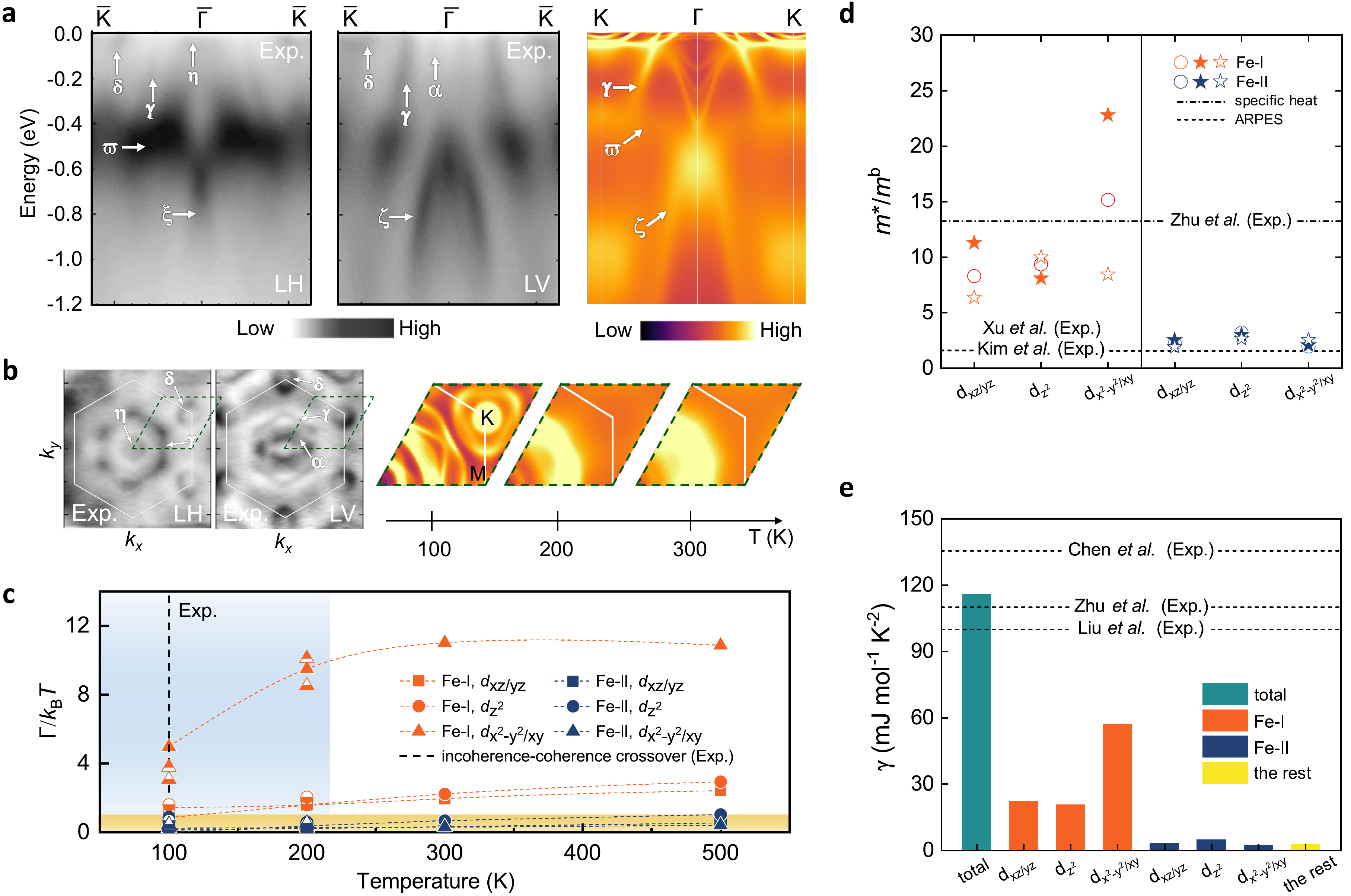}
	\caption{{\bf Comparison with reported experiments. }$\bf{a}$, The momentum-dependent spectral function $A(\mathbf{k},\omega)$ calculated from FM phase in comparison with ARPES data. The two figures on the left (expressed with black and white intensity) are the ARPES results measured with differently polarized photons of linear horizontal (LH) and linear vertical (LV) polarization taken from Ref.~\onlinecite{xu_signature_2020}. The rightmost figure is the $k_z$-integrated DFT+DMFT result at $T=100$~K. {It is found that each band has the significantly mixed characters of Fe-I, Fe-II and other anions.} $\bf{b}$, Fermi surfaces in the $k_x$--$k_y$ plane. The two left plots are the ARPES results (black and white) \cite{xu_signature_2020}. The right three plots (colored) show the $k_z$-integrated DFT+DMFT results: FM at $T=100$ and 200~K, and paramagnetic at $T=300$~K. ARPES data in ({\bf a}) and ({\bf b}) are presumably measured below 25~K \cite{xu_signature_2020}. $\bf{c}$, The calculated $\Gamma/k_{\rm B}T$ as a function of temperature. A vertical dotted line indicates the experimentally reported incoherence-coherence crossover temperature \cite{zhang_emergence_2018,corasaniti_electronic_2020}, and the colored areas highlight the regions for which $\Gamma \leq k_{\rm B}T$\cite{mravlje_coherence-incoherence_2011} (yellow) and $T \leq T_\mathrm{c}$ (skyblue). The upper- and the lower-filled symbols refer to the values of the majority and minority spin, respectively in the FM phase, and the fully-filled symbols represent the values of the paramagnetic phase. $\bf{d}$, The calculated $m^*/m^\mathrm{b}$ of each orbital at $T=100$~K for Fe-I (orange) and Fe-II (blue). The filled and empty stars indicate the values of the majority and minority spin, respectively in the FM phase, and the circles represent the values from paramagnetic phase. The horizontal dash-dotted and dashed line indicates the values from specific heat\cite{zhu_electronic_2016} and ARPES experiments\cite{kim_large_2018,xu_signature_2020}, respectively. $\bf{e}$, The calculated orbital-decomposed Sommerfeld coefficient in FM phase at $T=100$~K. Here `total' denotes the sum of all orbital contributions, and `the rest' the contributions from the other states than Fe $d$-orbitals. The experimentally measured values are represented by horizontal dashed lines \cite{chen_magnetic_2013,zhu_electronic_2016,liu_anomalous_2018}.		
		\label{fig3}}
\end{figure*}

{The new picture for $\mathrm{Fe_3GeTe_2}$ as a `site-differentiated Hund metal' provides useful insights to understand the experiments. Before discussing those features in detail, let us take a look at the spectral function $A(\mathbf{k},\omega)$ which can be directly compared with angle-resolved photoemission spectroscopy (ARPES) results, thereby enabling us to check the reliability of our parameter choices for interactions and the correlated electronic structure. Fig.~\ref{fig3}a shows the excellent agreement between theory and experiment.
For example, the pronounced spectral features, namely the bands assigned as $\gamma$, $\zeta$ and $\omega$ in Ref.~\onlinecite{xu_signature_2020} are well  identified in our $A(\mathbf{k},\omega)$. The near-Fermi-level states are also well reproduced; see, e.g., $\alpha,\delta$ and $\eta$ bands.}

The first experimental quantity we want to revisit is the large quasiparticle mass;  $m^*/m^\mathrm{b} = \gamma/\gamma^\mathrm{b} \simeq 13$ reported from specific heat measurement \cite{zhu_electronic_2016} (i.e., taken from the measured Sommerfeld coefficient $\gamma$ and its DFT estimate $\gamma^\mathrm{b}$). This value remains as a puzzle especially because ARPES data was deemed to be in fair agreement with band calculations $m^*/m^\mathrm{b} \simeq 1.6$ \cite{kim_large_2018,xu_signature_2020}. 
This large discrepancy has been highlighted throughout the literature \cite{zhu_electronic_2016,zhang_emergence_2018,xu_signature_2020}, but the issue remains still unresolved.

Here we argue that these seemingly contradictory experimental results can be understood as a natural manifestation of the site-differentiated Hund physics. In multiorbital systems, the mass enhancement measured by specific heat is mainly determined by the more correlated orbital \cite{de_medici_selective_2014}. In fact, our calculation shows $m^*/m^{\rm b} \approx 7$~--~23 for the more correlated site Fe-I, which is comparable to the value from specific heat experiment (dash-dotted line in Fig.~\ref{fig3}d).
It is due to that Sommerfeld coefficient $\gamma$ extracted from specific heat is a sum of each orbital contribution $\gamma_l \sim (m^*/m^\mathrm{b})_{l}$; see Supplementary Note~5 in SI for more details. As discussed in Ref.~\onlinecite{de_medici_selective_2014}, it is comparable to the series resistor in a circuit for which total resistance is given by the direct sum, and the total voltage can be approximated by the one applied to the large resistor. Indeed, the site- and orbital-decomposed values  (Fig.~\ref{fig3}e) clearly show that the Fe-I contributions  dominate the total value $\gamma_\mathrm{tot}$, whereas those of Fe-II are much smaller. The sum over all orbitals $\gamma_\mathrm{tot}$ eventually gives rise to good agreement with experiments.

The ARPES situation is quite different on the other hand. The mass enhancement of a given `band' is a weighted sum of each orbital contribution therein. In fact, the previous study revealed that most bands in this material have both Fe-I and Fe-II characters (see, e.g., Supplementary Fig.~3 of Ref.~\onlinecite{xu_signature_2020}). 
As inferred from $m^*/m^\mathrm{b} \approx 2$~--~4 of Fe-II being comparable to an ARPES estimate (Fig.~\ref{fig3}d), we presume that, at least in some bright bands captured by ARPES, the weaker correlated Fe-II $d$-orbitals as well as the other itinerant Ge and Te states possibly have the larger portion than Fe-I, which can result in the small $m^*/m^\mathrm{b}$ (see Supplementary Note~6 in SI for the related analysis).

Now let us examine the heavy fermion behavior and the incoherence-coherence crossover observed at $T \sim 100$~K \cite{zhang_emergence_2018,corasaniti_electronic_2020}. Figure~\ref{fig3}c presents $\Gamma/k_\mathrm{B}T$ as a function of temperature for both paramagnetic (fully filled symbols) and FM phases (partially filled). Here, the coherence temperature $T^*$ is defined by $\Gamma/k_\mathrm{B}T^*=1$ (see the yellow region in which $\Gamma \leq k_\mathrm{B}T$), namely the temperature below which quasiparticle lifetime exceeds the timescale of thermal fluctuation \cite{mravlje_coherence-incoherence_2011,ryee_2021}. We note that the sizable site-dependence of correlation makes electron scattering very different at two Fe sites: Whereas $\Gamma/k_\mathrm{B}T$ of Fe-II remains less than 1 even well above the experimental $T^*$, that of Fe-I is much greater. The larger contribution from Fe-I, particularly $d_{\rm x^2-y^2/xy}$ orbitals, and its significant temperature dependence as $T \rightarrow T^*$ indicate that the experimentally observed incoherence-coherence crossover around $T \sim 100$~K  \cite{zhang_emergence_2018,corasaniti_electronic_2020} is mainly attributed to the corresponding orbitals residing in Fe-I.

For comparison, we also investigate the isostructural material $\mathrm{Ni_3GeTe_2}$ whose $\gamma\sim$ 9 mJ/mol$\cdot\rm K^2$ is much smaller \cite{zhu_electronic_2016}. Our calculation reasonably well reproduces the experiment and the mass enhancements $m^*/m^\mathrm{b} \leq 1.5$ with no appreciable site-differentiation (see Supplementary Note~6 in SI for our calculation results).  This contrasting behavior, attributed to the absence of vHS and the different valence (see Supplementary Fig~6c in SI), renders $\mathrm{Fe_3GeTe_2}$ as a unique example of site-differentiated Hund metal.

Finally, another important issue for $\mathrm{Fe_3GeTe_2}$ is related to the nature of its magnetic moment. While in the early works it was understood or discussed within itinerant Stoner picture  \cite{zhu_electronic_2016,zhang_emergence_2018}, more recent studies emphasized the local nature of spin. 
For example, Xu {\it et al.} observed the nearly unchanged exchange splitting in their ARPES data  even up to $T > T_\mathrm{c}$ \cite{xu_signature_2020}. The situation is therefore reminiscent of a prototypical FM Hund metal SrRuO$_3$ whose exchange splitting persists above $T_\mathrm{c}$ due to the enhanced local spin character whereas the very low temperature behavior is well described within Stoner theory \cite{jeong_2013,kim_2015}. 
Note that, near $T_\mathrm{c} \simeq 220$~K, $\chi^\mathrm{spin}$ of Fe-I clearly indicates the local moment behavior (Fig.~\ref{fig2}c), and as a consequence, $\Gamma$ is large. As temperature decreases, on the other hand, $\Gamma$ gets reduced by the enhanced spin screenings, which eventually result in the long-lived quasiparticles well below $T \sim 100$~K (Fig.~\ref{fig3}c). Therefore our current study provides a unified picture for $\mathrm{Fe_3GeTe_2}$ within which its spin moment can be described as being localized above and itinerant well below $T_\mathrm{c}$.

%\vspace{3mm}\textbf{\\ Summary}$\;\;\;$

We demonstrated that $\mathrm{Fe_3GeTe_2}$ is a Hund metal with intriguing site dependence originated from the microscopic details in the underlying band structure. The proposed new picture of `site-differentiated Hund metal' not just makes this representative metallic vdW ferromagnet an even more exciting material platform, but also provides useful insights to understand the previous experiments including the ones that are seemingly contradictory to each other. Our results hopefully stimulate further experimental and theoretical investigations of the related systems. For example, the direct experimental observations of the site- or orbital-selective correlation effects can be an interesting research direction. 
%It can also provide the useful ground to manipulate thier electronic and magnetic properties via, {\it e.g.}, polarized photons \tcr{[Refs]}.

\section*{Methods}
\subsection*{Computation details of DFT+DMFT calculations}
The DFT electronic structure of $\mathrm{Fe_3GeTe_2}$ was obtained by using \textsc{FlapwMBPT} package \cite{kutepov_linearized_2017}, which is based on the full-potential linearized augmented plane wave plus local orbital method, employing the local density approximation (LDA). We used experimental lattice parameters as reported in Ref.~\onlinecite{deiseroth_fe3gete2_2006}. On top of this non-spin-polarized DFT band structure, 108 maximally localized Wannier functions were constructed with a wide energy window of [-15:10]~eV containing Fe-$s$, Fe-$p$, Fe-$d$, Ge-$s$, Ge-$p$, Ge-$d$, Te-$s$, Te-$p$, and Te-$d$ characters, and its tight-binding Hamiltonian was built through the \textsc{$J_X$} interface\cite{yoon_jx_2020}. We performed single-site dynamical mean-field theory (DMFT)\cite{georges_dynamical_1996,kotliar_electronic_2006} calculations for $d$-orbitals of Fe-I and Fe-II by employing \textsc{ComCTQMC}\cite{choi_comdmft_2019} implementation of 
the hybridization-expansion continuous-time quantum Monte Carlo (CTQMC) algorithm\cite{gull_continuous-time_2011} as an impurity solver. Namely, we solved two impurity problems, one for Fe-I and one for Fe-II, per DMFT self-consistency loop. Within our DFT+DMFT scheme, electronic self-energy is momentum-independent (i.e., local in space) \cite{georges_dynamical_1996}, and is assumed to be diagonal in cubic harmonics basis to avoid a sign problem by omitting the off-diagonal elements of hybridization functions whose values are actually small. 
The self-energy on the real frequency axis was obtained from analytic continuation of imaginary axis data using the maximum entropy method \cite{jarrell_1996}. For the magnetically ordered phase of \FGT, we used \textsc{EDMFTF} package \cite{haule_dynamical_2010} for DFT+DMFT calculations. 

For the double counting (DC) self-energy, we adopted the nominal DC scheme which reads $\Sigma_{\rm DC}(n_0)=U(n_0-\frac{1}{2})-\frac{J_H}{2}(n_0-1)$ \cite{pourovskii_self-consistency_2007}, where $n_0$ is the nominal charge. We took $n_0=6.0$ for both Fe-I and Fe-II, which is close to the $d$-orbital occupancy obtained from DFT ($n_{\rm DFT} \simeq 6.4$). Note also that this DC scheme with $n_0=6.0$ was used for this material in a previous study \cite{zhu_electronic_2016}, which yields good agreement of magnetic moment with experimental data.
While we mainly present the results obtained from $n_0=6.0$, we found that varying $n_0$ in a range of [5.4:6.6] does not lead to any qualitative changes; e.g., see Supplementary Note~7 of SI for spin and orbital susceptibilities obtained from different $\Sigma_{\rm DC}(n_0)$ values. 

The Hubbard $U$ and Hund coupling $J_{\rm H}$ of Fe-$d$ were chosen to be $U \equiv F^0=5.0$ eV and $J_{\rm H} \equiv (F^2+F^4)/14=0.9$ eV for both Fe-I and Fe-II, which were used in a previous study on $\mathrm{Fe_3GeTe_2}$ \cite{kim_large_2018}. These values are also comparable to the theoretically estimated values for iron chalcogenide family \cite{miyake_comparison_2010}. Here, $F^0$, $F^2$, and $F^4$ are Slater integrals and the ratio ${F^4}/{F^2}=0.625$ was used for parametrization of on-site Coulomb interaction tensor. For $\mathrm{Ni_3GeTe_2}$ which will be discussed in Supplementary Note~6 of SI, the same computation scheme was used except for the use of a larger nominal charge ($n_0=8.0$) for the DC self-energy and a larger Hubbard $U$ of $U\equiv F^0=6.0$ eV, considering its chemical environment on Ni sites. Note also that, for $\mathrm{Ni_3GeTe_2}$, we did not consider any magnetic order due to its absence even at very low temperatures \cite{deiseroth_fe3gete2_2006,zhu_electronic_2016}. We used experimental lattice parameters of $\mathrm{Ni_3GeTe_2}$ \cite{deiseroth_fe3gete2_2006}.

\nolinenumbers %%%%%%%%%%%%%%%%%%%%%%%%%%%  MJHMJH

\vspace{3mm}\textbf{\\ Data availability}$\;$
The data that support the findings of this study are available from the corresponding
author upon reasonable request.
	
\vspace{3mm}\textbf{\\Code availability}$\;$
All codes used in this work are accessible through their websites.

%\begin{Acknowledgements}

\vspace{3mm}\textbf{\\ Acknowledgements}$\;$ 
T.J.K. thanks Hongkee Yoon for discussion on \FGT at the initial stage and comments on the use of \textsc{$J_X$} interface for constructing tight-binding Hamiltonians. S.R. is grateful to Sangkook Choi for fruitful conversation.
This work was supported by the National Research Foundation of Korea (NRF) grant funded by the Korea government (MSIT) (Grant Nos. 2021R1A2C1009303 and NRF-2018M3D1A1058754), the KAIST Grand Challenge 30 Project (KC30) in 2021 funded by the Ministry of Science and ICT of Korea and KAIST (N11210105), and the National Supercomputing Center with supercomputing resources including technical support (KSC-2020-CRE-0084).

\vspace{3mm}\textbf{\\ Author contributions}$\;$
$^*$These authors contributed equally to this work; T.J.K. and S. R.. T.J.K. performed all the calculations and initiated the project by noticing the site-differentiation in $\mathrm{Fe_3GeTe_2}$. S.R. led the data analysis. M.J.H. supervised the project. All authors participated in the interpretation of the data and wrote the manuscript.

\vspace{3mm}\textbf{\\	Competing interests}$\;$
The authors declare no competing interests.\\[3mm]

\vspace{3mm}\textbf{\\	Additional information}$\;$
Correspondence and requests for materials should be addressed to M.J.H (mj.han@kaist.ac.kr).

$^\dagger$ Correspondence and requests for materials should be addressed to M.J.H. (email: mj.han@kaist.ac.kr).
%\end{Acknowledgements}

\newpage

\bibliography{FGT_bibtex}

\newpage

\end{document}